# Random moves equation Kolmogorov-1934. A unified approach for description of statistical phenomena of nature

G. S. Golitsyn

A.M. Obukhov Institute of Atmospheric Physics, Russian Academy of Sciences

*Introduction*

In 1934 A. N. Kolmogorov published a two-page paper [1] which is able to describe most macroworld probability phenomena and considered up to the present as empirical laws. In the first half of the last XX century it was the frequency-size distribution, the Gutenberg – Richter law. In 1941 Kolmogorov and Obukhov proposed turbulence laws on the grounds of the dimensional analysis. But at the time of mid XX century the geophysical science had not known many other empirical statistical laws.

In 1958 it was in Oxford, UK, an International Symposium on Turbulence and Atmospheric Pollution. A. M. Obukhov presented a paper "Description of turbulence in Lagrangian terms" [2]. Without referencing exactly to [1] he was describing three second probability moments of the basic equation of (1) for random moves (as the paper [1] was called: "Zufallige Bewegungen"). The corresponding equation for the probability density distribution function $p(t,u_i,x_i)$ for its time evolution under markovian forcing

$$\frac{\partial p}{\partial t} + u_i \frac{\partial p}{\partial x_i} = \frac{D}{2} \frac{\partial^2 p}{\partial u_i^2} \qquad (1)$$

is having a new random variable as velocity. This equation differs from random walks described by the Fokker – Plank equation. The fundamental solution of this equation was first presented by Monin and Yaglom (1975 [3] in their §24)

$$p(t,u_i,x_i) = \left(\frac{\sqrt{3}}{2\pi Dt}\right)^3 \exp\left[-\left(\frac{u_i^2}{Dt} - \frac{3u_i x_i}{Dt^2} + \frac{3x_i^2}{Dt^3}\right)\right].$$

A. M. Obukhov presented three second moments of the probability density distributions $p(u_i,x_i)$ as

$$<u_i^2(t)> = C_1 \varepsilon t \equiv \bar{u}^2, \qquad (2)$$

$$<u_i x_i(t)> = C_2 \varepsilon t^2 \equiv K, \qquad (3)$$

$$<x_i^2(t)> = C_3 \varepsilon t^3 \equiv S \approx r^2, \qquad (4)$$

with $\varepsilon = 2D$, where $C_i$ are non-dimensional constants should be found at comparison these moments with empirical data. As quoted by Bridgman [4], A. Einstein has warned [4] that for reasonable results constant $C_i$ should not be very large or very small, i. e. $C = O(1)$. A review of such constants for turbulence can be found in [3, 5] and they are $C_1 \approx 1.6$, $C_2 \approx 0.2$ and $C_3 = O(10)$. In [2] $\varepsilon$ is the rate of change of the kinetic energy for turbulent motions. We assume that (2) is the mean square of velocity, (3) is the coefficient of turbulence exchange, and (4) is the area of the random moves. To declare these one needs some kind of ergodic hypotheses, and we accept it because it produces reasonable results for natural process [5]. Obtaining from (4) as was done in [2]

$$t \equiv \tau = (r^2 / \varepsilon)^{1/3} \qquad (5)$$

and substituting it into eqs. (2) and (3) we obtain Kolmogorov – Obukhov's results of 1941 and the Richardson eddy-diffusion law (see also [5]).

One should recall here a very important notion of intermediate asymptotics introduced by Barenblatt and Zeldovich [6] determining the range of scales for which the obtained results are valid. This can be clarified at comparison of (2) and (4) with empirics. For this we must analyze the dimensions of the parameters governing the process or the phenomenon and see whether these parameters can form a non-dimensional similarity criterion. In the case of turbulence it is the Reynolds number $\text{Re} = uL/\nu$ with $L$ a large scale of the flow, $\nu$ is the kinematic viscosity. So for turbulence of the 1941 the intermediate range is the inertial interval and the very weak dependence of $C_i$ on Re could be noticed [7]. Most of the results are in my book [5] published in Russian in limited number but here there are some new results with tornadoes, see the description of this topic later for earthquakes here. We want also to specify here in plain words that we assume here that $\varepsilon$ is the rate of generation of any kind of energy, stress or potential in consideration.

*Immediate results from the inspection of second probability moments*

Several important results can be obtained from (4) by close inspection of the second moments (2-4). First it was shown by A. M. Obukhov in 1958 at Oxford International Symposium on Turbulence and Air Pollution [2]. He demonstrated there, that time $\tau$ as in (5), extracted from (4) and substituted into (1) and produced $\langle u^2(t) \rangle = (r^2/\varepsilon)^{1/3} \varepsilon = (\varepsilon r)^{2/3}$, the result of 1941. Substituting this time into (3) he got the Richardson – Obukhov law of 1926, 1941 [3], as was shown above.

One may consider (2) as the velocity structure function with zero or small initial condition and knowing the mutual relations between spectra and structure functions [3] one obtains $S_u(k) = \varepsilon^{2/3} k^{-5/3}$ [2]. In 2010 Gledzer and Golitsyn [8] numerically tested how well these moments are fulfilling in dependence on the size of the studied ensemble, i. e. on the number *N* of calculated particles. The results are presented at the Fig. 1. It is seen that at *N* = 10 already the mutual relative velocities between $N(N-1)/2 = 45$ relative distances between the particles are observed rather well. The calculated distances between the particles are obtained by integration of their velocities increased with time as $t^3$ almost ideally, because the integration is an additional smoothing operation. All probability theory formulas are for $N \to \infty$, however our work [8] demonstrates that this infinite size ensemble asymptotes as started to be approached rather early. Fig. 1 shows that at *N* = 100 both asymptotes coincide with the theoretical one. The eddy diffusion coefficient (3) is started to be approached [8] only at $N \geq 300$.

In [8] we demonstrate that the moments (2-4) are fulfilled also for one-dimensional movements. Because wind waves on the sea are mostly vertical, then their time spectrum should $\omega^{-4}$, at least its main energy containing part (see [5]).

The second moment (4) explains readily several phenomena area related. The most challenging was cumulative distribution of plates [9]: $N(\geq S) = 7S^{-n}$, $n = 0.33$, which was obtained for sizes from 0.002 to 1 steradian disregarding 4 small plates down to $10^{-4}$ str and 7 large continental plates $S > 1$ str. Because the cumulative distributions have dimension of frequency, i. e. inverse time and from (4) it should be $N(\geq S) \approx \left(\dfrac{C_3 \varepsilon}{S}\right)^{1/3}$, or from probability theory $n = 1/3$. The final explanation [10-12] was used $\varepsilon \approx 1 \cdot 10^{-11} \text{m}^2\text{s}^{-3}$ = 406 str/s³. From moment (4) and from the two above distributions we see that $C_3 = 7\varepsilon^{-1/3}$ wherefrom $C_3 = 0.945 \approx 1$ which perfectly satisfies the warning by Albert Einstein (see [4], ch. 8). Similar distribution was obtained for masses of galaxies close to our Milky Way at the assumption that their masses are proportional to their disc areas [5].

The dimension of the second moment $\langle u^2(t) \rangle$ coincides with the energy per unit mass. The energy in the nature is conserved but may transform between kinetic, potential, stress and so on. This opens the road of using eqs. (1-4) to any kind of processes if we may determine their forcings denoted above as $\varepsilon$, the rate of generation of the energy in study. This we demonstrate for the earthquakes.

The equation (1) is for parabolic type and has several interesting properties. Under the transformation of its independent variables as it can be reduced to non-dimensional form

$$\frac{\partial p}{\partial \tilde{t}} + \tilde{u}_i \frac{\partial p}{\partial \tilde{x}_i} = \frac{\partial^2 p}{\partial \tilde{u}_i^2}, \qquad (6)$$

where $\tilde{t} = t/\tau$, ($\tau$ see in (5)), $\tilde{u}_i = u(\varepsilon t)^{-1/2}$, $\tilde{x}_i = x(\varepsilon t^3)^{-1/2}$ – are non-dimensional. The moment for velocity (2) $<u^2(t)> = \varepsilon t$ – can be considered as energy moment for the unit mass, and $<x_i^2(t)> = S$ as the area of moves within which the ensemble of considered particles can move. The product of (2) and (4) describes the area over which the energy is contained and the volume $Sh\bar{u}^2$, $h$ is the height, describes the energy of the whole observed process

$$E = \rho h \bar{u}^2 S. \qquad (7)$$

The decade of 1990s was called by UN the decade of combating the natural catastrophes. A governing scientific Council was established with Sir James Lighthill as a chairman. The most damaging are hydrometeorological phenomena and earthquakes, the last ones are fortunately much more seldom. At a meeting of this Committee Lighthill in 1995 said that until that time we could not understand how tropical hurricanes can reach energies up to thousands megaton bombs of their kinetic energy. This was shown [13] for hurricanes with forcings $b = \varepsilon$ in [5], $h \sim 18$ km when [11] mass of air is $M = 10^4$ kg/m² and Coriolis term equals to $l_c = 2\omega \sin\theta = \frac{4\pi}{T}\sin\theta$ we obtain from (7) transformed to (this term vanishes near equator)

$$E = Mb^2 l_c \qquad (8)$$

and substituting there, see [13], $b = 3 \cdot 10^{-2}$ m²/s³, $\theta = 20°$ we obtain $E \approx 10^{19}$ J, with trotil equivalent 1000 kg = $2.7 \cdot 10^{12}$ J. We obtain the needed power.

In the same way we can estimate destructing power of tornados and similar vortices when there are data with their diameters, $d$, and mean velocities $\bar{u}$ as Fujita scales (the results are in press). Diameters produce the area of vortices $S = \pi r^2$, $r = d/2$ and from the moment (4) we have $S = C_3 \varepsilon t^3$, from the moment (2) $\bar{u}^2 = C_1 \varepsilon t$, the paper [40] we estimate time associated with vortices as $t = \left(S/\bar{u}^2\right)^{1/2}$. In the above paper we consider over 150 vortices recorded in the RF at 2001 – 2021 with diameters from 20 to 1000 m and velocities from 25 to 105 m/s. We estimate their forcings as $\varepsilon = \bar{u}^2/t = \bar{u}^3/S^{1/2}$. The destructing power estimated from Fujita index were calculated for a volume 10 m · 100 m² = $10^3$ m³, the characteristic size of suburban dwellings, and found to be from $10^7$ to $10^{12}$ J, or in our age of terrorism from 0.5 kg to several tons of TNT (half a kilo of TNT is enough to blow up a car). Forcing is the energy released in one second.

*Earthquakes*

The non-dimensional form of the eq. (6) allows us to use for arbitrary processes describing any form of the energy transformation. Let us turn to the earthquakes that are observed near boundaries of the lithospheric plates. The thickness of these plates is of order $L = 30$ km. Below them is the convecting mantle [12]. The convection is not uniform and plates move with velocities of several cm/yr and plates interacting with each other are generating stresses in both of them. The stresses are realized producing EQs. The rate of the seismic energy generation can be estimated from the size-frequency distribution of EQs over a certain period [11]. This estimate is evidently from below but it still may produce reasonable results. This estimate produces $\varepsilon = 1 \cdot 10^{-11}$ m²/s³.

Let us return to the Gutenberg – Richter, GR, law for the EQs. In my first paper on quakes [14] I have based on the knowledge that in the usual form of the EQs

$$\lg N(\leq m) = a - bm, \ b \approx 1, \tag{9}$$

where $m$ is called magnitude. Later on (see [15]) the seismic moment has been introduced as a measure of EQ energy

$$M = \mu S u, \tag{10}$$

where $\mu$ is the Young modulus of the earth core material, $S$ – the area of the EQ rupture, $u$ – the mean shift of the core blocks along the rupture. The seismic moment and the magnitude are related as [15]

$$m = \frac{2}{3}\lg M - 6. \tag{11}$$

In the terms of $M$ the GR law is presented as

$$N(\geq M) \sim M^{-2/3}. \tag{12}$$

The relation (11) statistically is fulfilled with the determination coefficient $r^2 \approx 0.8$ (see [16]). This author was introduced into seismology in 1995 by Y. Y. Kagan while being in JPL for a week on climate change issue. Kagan presented me with news in seismicity, especially by the paper [17]. In this important paper it was discovered that the value $b$ in GR law (9) increases to about 1.5, for seismic moment $M \geq 1.6 \cdot 10^{20}$ H·m and also for EQs near midoceanic ranges producing new core. Before that Kanamori and Anderson [18] paid an attention to the released stress which changes in rather narrow limits $30 < \Delta\sigma < 70$ bar (1 bar = 1 atm = 0.1MPa = $10^5$ H·m with median value 4.4 MPa [5]). The paper [18] explains several empirical seismology relations.

Now we are ready to develop a similarity theory for EQs [14, 5]. As determining parameters for the process of EQs we use the material constants of the core material [15]. The Young modulus $\mu = (3 \div 7) \cdot 10^{10}$ N/m², density $\rho = 3 \cdot 10^3$ kg/m³ and the value of the released stress $\Delta\sigma \approx 4$MPa $= 4 \cdot 10^6$ N/m². The value $\Delta\sigma / \mu \approx 10^{-4}$ and this small non-dimensional parameter we shall not further consider as changing little. As was shown in [17, 14] the plate thickness $h$ is also important.

All geodynamic processes are caused by geothermal heat flux which is equal globally to $F = 4.5 \cdot 10^{13}$ W at mean density 86 mW/m² [10]. From the above values we construct the scales of length and time

$$L = (M/\Delta\sigma)^{1/3}, \qquad (13)$$

$$T = M/F, \qquad (14)$$

which are characteristic for individual quakes. The seismic moments $M$ are changing at the registered quakes over 16 orders of magnitude [5, 15]. We may form one non-dimensional similarity criterium as

$$\Pi = \frac{L}{h} = \frac{M^{1/3}}{\Delta\sigma^{1/3}h} \qquad (15)$$

as $M = M_{cr} = 1.6 \cdot 10^{20}$ H·m, $\Delta\sigma = 4.4 \cdot 10^{6}$ H/m², $F \approx 4.5 \cdot 10^{13}$ W (kg·m²s⁻³) and $h \approx 30$ km we obtain $\Pi = 1.07 \approx 1$ (see [16]). This value $\Pi \approx 1$ separates the EQ spectral intervals, or GR law, with $b \approx 1$ at $\Pi > 1$ at $m \leq 7.5$ from $\Pi < 1$ and with $b \approx 1.5$ when EQs are one and half more seldom and then $N(\geq M) \sim M^{-1}$. The unified formula for quakes may be written as

$$N(\geq M) = FM^{-1}f(\Pi). \qquad (16)$$

Data of global catalogs [19, 17] in dependence on the moment $M$ are presented for 828 events at 1977 – 1993. This data has been recalculated in $\Pi$, as the similarity criterium for each EQ.

For $\Pi < 1$ the function $f(\Pi)$ in (16) was decomposed into the McLoren series and it should start with linear term (no forcing – no quake)

$$f(\Pi) = C_1\Pi = C_1M^{1/3}/\Delta\sigma^{1/3}h \text{ and } N(\geq M) \sim C_1M^{-2/3}, \ C_1 = 0.35. \qquad (17)$$

For the case $\Pi > 1$ it was found that $f(\Pi)$ is about const $= C_2 \approx 0.34 \pm 0.02$ [16]. Both coefficients $C_1$ and $C_2$ are close to 0.35 at $\Pi \to 1$.

The same data give a possibility to estimate numerical coefficients at the EQ parameters: time $\tau = L/C$, $C = (\mu/\rho)^{1/2}$ the scale is velocity of volume and surface seismic waves for each EQ:

$$\tau = 14L/C, \quad L_i = 2.3L(M_i/\Delta\sigma)^{1/3}, \quad S_i = 0.34L_i^2 = 0.34(M_i/\Delta\sigma)^{2/3}$$

and the shift of adjacent blocks along the quake disruption

$$u = 0.54(M/S\mu) = 0.54M^{1/3}\Delta\sigma^{1/3}\mu^{-1},$$

wherefrom even for the Chile 1960 event the shift at the neighboring blocks $u$ is only about 20 m [16].

There are quakes after the human activity: due to filling up large water reservoirs after extracting large volumes of oil and gas etc. The quakes are not large with $m \leq 4$ and seismic moments $M \leq 10^{15}$ H·m. Statistics of such event are described satisfactorily by the Gutenberg – Richter law [21]. At that time the author did not know total capacity of Kolmogorov (1) shortened version $\partial p/\partial t = D\partial^2 p/\partial u^2$. This produce $\langle u^2(t)\rangle = Dt$, but coefficient $D$ do not contain any information about forcing. The forcing for these events is the time change of internal pressure in the core disturbing the existing isostatical equilibrium, the fracture of the material releases the stress and we feel it as induced EQs. The internal pressure vertical gradient is

$$\frac{dp}{dz} = -\rho g.$$

Time change of this term, dimension is $ML^{-1}T^{-3}$, causes the change in the already existing stress in so called isostatic equilibrium and the new stresses are released by new quakes (see [5], §3).

In [5] there are data on statistics of medium size tsunamis at the RF Far East. In the range of wave heights from centimeters to a meter the cumulative frequency of such events is inversely proportional to their heights, i. e. to their energy (see wind sea waves at [5], §6). These waves have been generated by underwater quakes near midoceanic ridges where the earth core is new and thin. Such observations demonstrate again the connection of quakes with ANK34. For the discrete processes one should expect $\langle u_i^2(t)\rangle = \varepsilon t_i$ and $N(\geq M) \approx C\varepsilon_g/M$ (see above, $C \approx 0.35$).

We present Fig. 2 as the direct confirmation of the Kolmogorov suggestion on the markovian ($\delta$-correlated) forcing at the presentation of eq. (1). This is the spectrum of microseisms in the period from 0.1 to 2400 seconds [22]. There are two regions proportional to $t^4 \sim \omega^{-4}$ showing microseism acceleration as white noise, so velocities are $\sim \omega^{-2}$ and displacements are $\sim \omega^{-4}$, or $t^4$. The first interval is due to large sea waves as was explained by Longuet-Higgins [23] and the second interval in minutes is due to human surface activity: transport, construction work etc., as I was explained by highly experienced seismologists. For this author Fig. 2 is the most bright illustration of the white noise origin of accelerations [23] when velocities $\sim \omega^{-2}$ and the displacements $\sim \omega^{-4} = t^4$.

The same statistical behavior reveals the specific starquakes [24], producing soft γ-rays periodically similar to GR law [24, 5]. We just mention this fact which illustrates the fulfillment of the eq. (1) laws also at the universal scales. The fact of discovering of such γ-ray repeaters is the sequence of the cold war when the owners of nuclear bombs agreed not to continue the tests of such weapons in the air, water, and space. And, as a control for the space, special satellites were issued. And it took some time to find out that signals are coming from space [5]. The most intense explosion was registered at 24.12.2004 (Wikipedia) with about $1.5 \cdot 10^{39}$ J of energy released (see [5]). This is 3-4 orders of magnitude less than at the bursts of supernova stars, but registered not only by satellites but also by the surface network. The explosions are processes on neutron-stars of 10 km size. The theory used to describe the processes estimates the time of explosion as of order $10^{-1}$ second and at that date the event was registered at satellites and at the surface during 100-200 ms, demonstrating that science may understand processes at far ends of our Galaxy – Milky Way.

*Energy spectrum of cosmic rays*

This natural phenomenon was awaiting its explanation and statistical description during more than half a century. The cosmic rays, CR, have been discovered only in 1912 by austrian physicist W. Hess who found that air ionization is increased with heights. The ionization plays a role in condensation processes so it is important for climate changes. At Fig. 3 there is the integral (cumulative) spectrum of the CR. The CR are produced at the explosion of superstars which in our Galaxy happen 2-3 times a century. The book by V. L. Ginzburg [25] describing the processes of CR birth is of interest up to now. Fig. 3 presents a weak decline at the energy about $3 \cdot 10^{15}$ eV due to the finite size of our Galaxy – Milky Way [7]. The basic idea on the acceleration of CR particles belongs to E. Fermi, who proposed the inhomogeneities of the galactic magnetic field. The numerical formulation of this idea was presented as the acceleration of CR particles at the fronts of collisionless shock waves [26]. The energy balance estimates produce [25] for the volume density of CR $w \approx 0.5$ eV/cm$^3$. The same is the estimate of the magnetic field $H = 5 \cdot 10^{-6}$ Gs and its energy $H^2/8\pi \approx 10^{-13}$ J/m$^3$.

The energy spectrum is calculated from the measurement as the number of particles with energies $E \pm dE$ counted per unit time per unit area and also coming from a unit space angle (all space is $4\pi$ steradians). The theoretical deviation of the forms of this spectrum was described only in 1997 [27] by dimensional arguments. For the system of measurements units we accept time $T$, surface $S$ and energy $E$ (instead of $mc^2$). For finding the integral spectrum $[I(\geq E)] = S^{-1}T^{-1}$ we also use volume density of CR $[w_0] = ES^{-3/2}$ and the rate of generation $G$ released at superstar explosions, $[G] = ET^{-1}$ and the energy itself $[E] = E$. From these parameters $I(\geq E) = G^{-i} w_0^{-k} E^l$ we find that $i = -1$, $k = 2/3$, $l = -5/3$, or

$$I(\geq E) = a_1 \frac{G}{E}\left(\frac{w_0}{E}\right)^{2/3} \sim E^{-5/3}, \tag{18}$$

where $a_1$ is a non-dimensional numerical coefficient of order $10^{-27}$. If we take as the unit area the square of the thickness of galactic disc 200 parsec $\approx 6 \cdot 10^{18}$ m, then the area unit will be $\sim 4 \cdot 10^{37}$ m$^2$ and $a_1 \approx O(1)$ (remember Albert Einstein [4]!).

The most accurate long, several years, measurements of CR particles are described in [28] and they have produced the energy exponent for the differential spectrum

$$I(\geq E) = \int_E^\infty I(\geq E) dE \sim E^{-n}, n = 2.67 \pm 0.02 \tag{19}$$

which means $I(\geq E) \sim E^{-n+1} \sim E^{-1.67}$. The integral spectra reduce fluctuations of differential spectra and we should expect that $n - 1 \approx 1.67 \pm 0.01 \approx 5/3$.

In the book of Ginzburg [25] one can find that after the "knee", or $E > 10^{16}$ eV the exponent $-n + 1 = 2.1$, or $2 + 0.1$. In [29] Golitsyn obtained for the integral spectrum the exponent $-n+1 = 19/9 = 2+1/9$ assuming that after $3 \cdot 10^{15}\,\text{eV} = 3 \cdot 10^{16}\,\text{GeV}$ the source of CR particles is decomposed by those whose Larmor radius is feeling the thickness of the disk and leaving the inside galaxy accelerations. More detailed consideration of the processes and their connection to eq. (1) can be found at the end of §4 of the book [5]. The equation of motion of CR particles is

$$\frac{dp}{dt} = f,$$

or the derivative of a momentum is equal to acting force, or acceleration. The surface physically does not enter into the accelerations and it is only due to our technical use of the determination of the spectrum. In other cases as for the earthquakes, hydraulics and destruction processes occurring through the surfaces, the surfaces must be considered. Here the unit surface is a part of the measurement procedure.

*Surface phenomena and relief*

Surface of bodies is formed by slopes along which water is flowing, the material is falling, they resist to winds etc. So the forcing is $g\sin\theta$ and the angle $0 < \theta < \pi$ being random. So the energy of the processes along the slopes is time proportional. In [30] one finds several statistical characteristics of the relief spectra. There are 24 spectra for the range $2 < \lambda_y < 60\,\text{km}$ measured at the state of Oregon for plain, hilly and mountain regions and all them are power form $k^{-n}$, $n = 2.03 \pm 0.04$.

The derivative of the slope against horizontal is the slope angle and the spectrum of this angles $\varsigma$ at $n = 2$ is equal to

$$S_\varsigma(k) = k^2 S(k) = D_1 / \pi = \text{const}, \tag{20}$$

i. e. the white noise spectrum, or of a markovian process or

$$\langle \varsigma(y_1)\varsigma(y_2) \rangle = \varsigma^2 S(y_1 - y_2). \tag{21}$$

In the book [30] the exponent $n = 2$ was related to Hausdorf measure but no its connection to slopes is mentioned. The sky bodies are finite sizes and their spectral harmonics are discrete. The harmonic analisis of the satellite velocities and their surface change measurements soon revealed the time fluctuations of such signals. It was noted that amplitudes spherical harmonics numbered *n* decrease as $n^{-2}$. It was called the Kaula rule [31, 32], who first presented this result. The authors of [32] fulfilled the whole spherical harmonical analisis of the data with stochastic forcing and found that the amplitudes of *n*-th harmonics decreased as $[n(n+1)]^{-2}$, slightly faster than $n^{-2}$.

Authors of [33] has extended such analisis up to $n = 43200$ for Earth, Moon and Mars where $n = 21600$. At their graph the $[n(n+1)]^{-1}$ dependencies are observed well for the Moon down to about 4 km and for Earth and Mars down to 1 km. Farther on they are approaching to $n^{-4}$. This was explained in [34] as the appearance at short distances of a correlation between slope angles [35] in the simplest form as

$$B_\theta(\theta(y_1)\theta(y_2)) = \exp(-\beta y) \qquad (22)$$

where $\beta = 1/y_0$ is the inverse wave length at which such a correlation is appearing. In [34] it was proposed that $\beta = g/c_0^2$, $c_0^2 = \mu/\rho$, $\mu$ is the Young modulus. On the Moon the gravity is six times smaller than here and this effect appears earlier.

Therefore the long time unexplained Kaula's rule $n^{-2}$ is the sequence of the Kolmogorov-34 results [1] interpreted properly [35]. The other consequences are fractal dependencies observed in hydrology [30, 5]. For instance the cumulative length distribution for rivers is $N(\geq l) \sim l^{-n}$, $n = 1.9$ and for the lakes area distribution is $N(\geq S) \sim S^{-n}$, $n = 0.95$, or for their mean size distribution is also 1.9. The exponent 1.9 is slightly less than 2 and $0.95 < 1$ that is showing a role of other processes, mostly climatic ones, determining their sizes. The slopes are governing flows therefore their cumulative distributions should be described by (1) and then $n = 1$, but other processes may slightly decrease it. So one may consider 1 as a necessary condition for action of the random move eq. (1).

At Fig. 4 there is a cumulative distribution of a number of floods in dependence on the damage from them [36, 37]. The dependence is a power one with the exponent 0.65 and the determination coefficient by eye is about $r^2 \approx 0.75$. I have during 20 years this plot before I have started to work on the paper [37]. At that time the paper [38] was published for cumulative distributions of world rivers with the clarifying results

$$P = cA^b, \quad (23)$$

$$N(\geq P) = aP^{-\beta}, \quad (24)$$

where $P$ is the area of the sea surface mudding mushroom from a river, $A$ – the watershed area of a river, $a$, $c$ – the numerical coefficients. The exponents are $b \approx 0.6 \div 0.7$ and $\beta = 1.02 \pm 0.03$. The value of $b$ is covering 2/3, and $b \approx 1$ close to the above. The exponent 2/3 appears evidently from the volume of precipitation over of a watershed but the sign of revealing extrawater shows up is by the surface flow. In other words it is a sign of a surface phenomenon caused by a volume effects. We have seen it already at the statistics of earthquakes most of which are due to the surface cracks within the earth core and therefore in the Gutenberg – Richter law $b$ is close to 2/3. For the cosmic rays spectra are measured technically at a unit surface but their acceleration is the volume process which is the cause to appearance thirds in their spectra exponents. The equation (1) for the random moves is a reach source for supplying both types of exponents: one and with thirds.

In the book [5] one could find other examples of fractal distributions. We want to add the mass distribution of stones from 1 to 2000 kg on the surface of Mars [39]. Four such examples are power laws $N(\geq m) \sim m^{-n}$ and the exponents for them are 0.9, 0.9. 1.1 and 1.2. These stones are formed by cosmic bodies targeting Mars and confirm the general form $N(\geq A) \sim A^{-1}$. Small deviations of the $n$ exponents from unity may be understood as a unsuffient volume of samples, but the general form is in accordance with eqs. (2) and (23).

*Discussion*

From the way of computation of eq. (1) in [8] it becomes clear that it is the statistical generalization of the both dynamical laws by Isaak Newton for an ensemble with $N$ particles. The theory of probability is presize for $N \to \infty$, but here we show that it starts to work well even for $N \geq 10$.

The computation of the second moments (2) – (4) is by averaging markovian accelerated signals within an ensemble of $N$ particles first by obtaining their velocities and then by second dynamical law obtaining coordinates of the particles. In the first step we compute the mutual velocities between total number $N(N-1)/2$ such pairs and average, and then mutual distances between such the pairs. The squares of velocity have the dimension of the energy and the square of mutual distances is of area dimension [8].

The velocity square (2) is time proportional so its frequency spectrum is proportional to $\omega^{-2}$, the area spectrum is proportional to $\omega^{-4}$. The first moment (2) corresponds to many random processes with the energy transformations, the moment (4) corresponds to frequency spectrum of wind sea waves [3 или 5?]. Numerical coefficients $C_i$ at the moments may in its turn depend on the values of the similarity parameters П and comparing theory with experiment one always should remember on that [7, 14, 5]. Here it happens to be in earthquakes and hydrology. The expansion decomposition of $C_i$ to the Taylor series up to linear term on или or П may quickly explain the empirical regularities.

Here, and in [5], we demonstrate that the law of random moves [1] by A. N. Kolmogorov describes well the probability distributions of various processes in nature. Combined with the dimensional arguments taking into account the nature and physics of the processes in consideration it explains the deviations with pure statistics as for earthquakes, hydrology processes and so on. At the very end it is proper to present a remarkable words of Andrey Nikolaevich Kolmogorov (cited by G. I. Barenblatt in [7]): "Stochasticity is a necessary if not the most important element of the universe, but it has a certain order in itself leading to concrete stable structures. The stability is limited in space and time which are specific in concrete situations". So this almost a century old paper by Kolmogorov may be used as an illustration of a final aim of the science in the sense of J. W. Gibbs who said: "The aim of science is to find a view from which the problem can be solved most simply and naturally", at least for the science of our environmental world. These two sayings are used as epigraphs for the book [5].

*Acknowledgements*

I started to work in science in 1958 under A. M. Obukhov, A. S. Monin, A. M. Yaglom and G. I. Barenblatt in turbulence and similarity theory. After about forty years

I have started to look around at scientific disciplines and for geodynamics I was introduced by Y. Y. Kagan and V. F. Pisarenko, for wind sea waves by Yu. I. Troitskaya and I have many others for discussions of various disciplines here and in US and Europe. Many grateful thanks for those colleagues and discussants who brought to me new and unusual topics like $\tau^4$ for seismic noise in 1997 and $S^{-1/3}$ for lithospheric plates distributions in 2004. A. N. Kolmogorov and V. P. Maslov were telling me how it is useful and necessary to look around in science.

*Appendix*

It is worth to present the main steps of the computations in [8] conducted by E. B. Gledzer for an ensemble by *N* terms. For each term the acceleration is prescribed which is markovian in time and with probability distribution among the terms as $\beta$, $\gamma$ and normal with the well known dispersions *D*. Time step $\Delta t$ determines the value of forcing, i. e. $\varepsilon = D\Delta t$. The choice of the probability distribution function for the random accelerations and the value of the time step does not influence the form of the results.

Most of computations were performed at $\Delta t = 0.01$. In [8] we consider the dynamical equations

$$\dot{u}_i = a_i, \quad \dot{x}_i = u_i \tag{П1}$$

with $i = 1, 2. \ldots, N$ – the general number of couple of such equations. Markovian probability distributions, or delta like time accelerations for each term $N_i$ are modelling energy introduction into system of the terms. In these conditions at $N \to \infty$ for the second moments of velocities and distances between the particles the following formulas are valid for the second moments [8]:

$$\langle u^2(t) \rangle = \varepsilon t, \tag{П2}$$

$$\langle x^2(t) \rangle = \frac{1}{3}\varepsilon t^3, \tag{П3}$$

compare with the fundamental solution (2). In the process of computation we compute the relative velocities between $N(N-1)/2$ particles, then the relative distances between them and average the results which are presented at Fig. 1. The last two results at the figure therefore present the average behavior within the system of $N$ particles obtained after $3.0/0.01 = 3000$ computed time steps. My role was in starting the computations and in the choice of coordinates for transformation of eq. (1) to non-dimensional form, also in interpretation of (2) as an energy for the unit mass. So this equation describes the time growths, say, of elastic energy before the earthquake relieves it.

This is the best illustration to the epigraph of Andrey Nikolaevich.

The list of figures for the paper by G. S. Golitsyn

Random moves equation by Kolmogorov 1934.

Fig. 1. The time behavior of second probability moments of Kolmogorov eq. (1).

Fig. 2. The spectrum of microseisms from [22].

Fig. 3. The integral cumulative energy spectrum of cosmic rays from Wikipedia.

Fig. 4. Cumulative distribution of the number of floods in dependence on their damage.

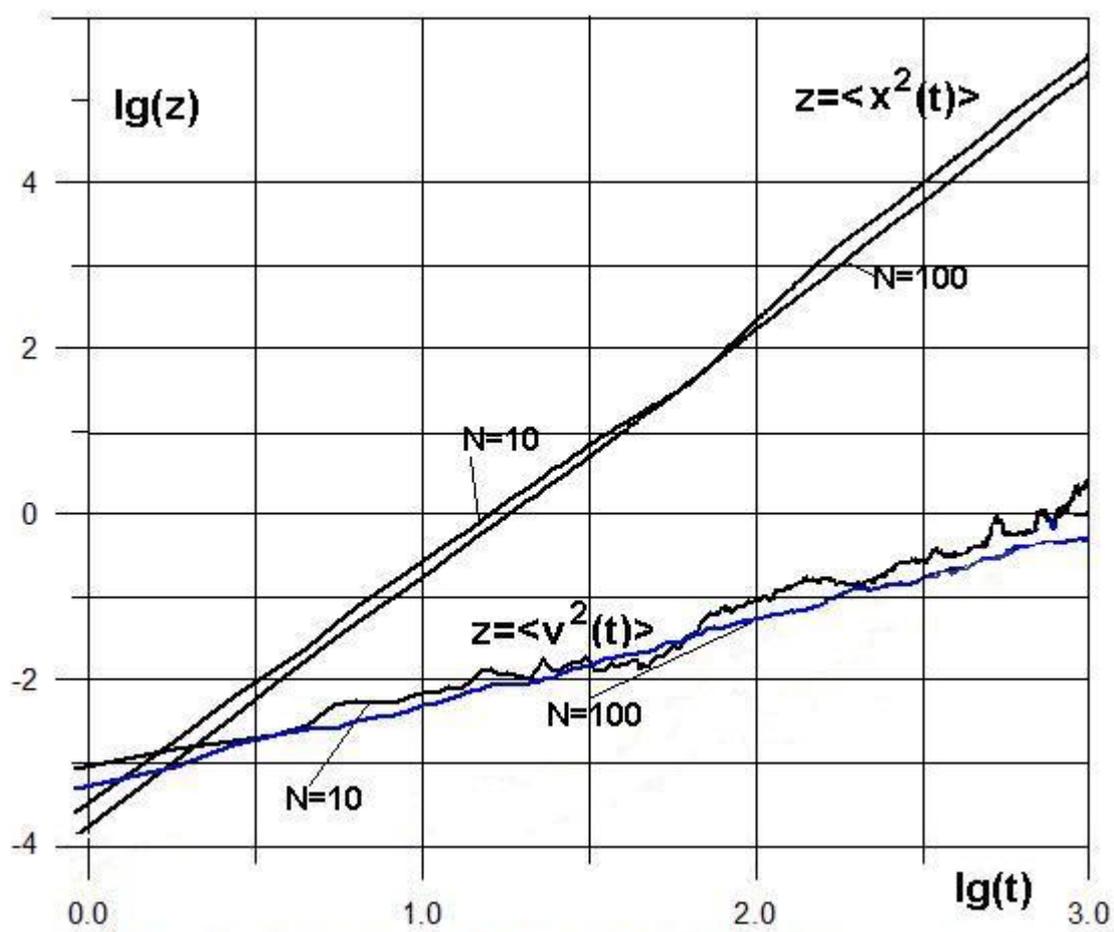

Fig. 1.

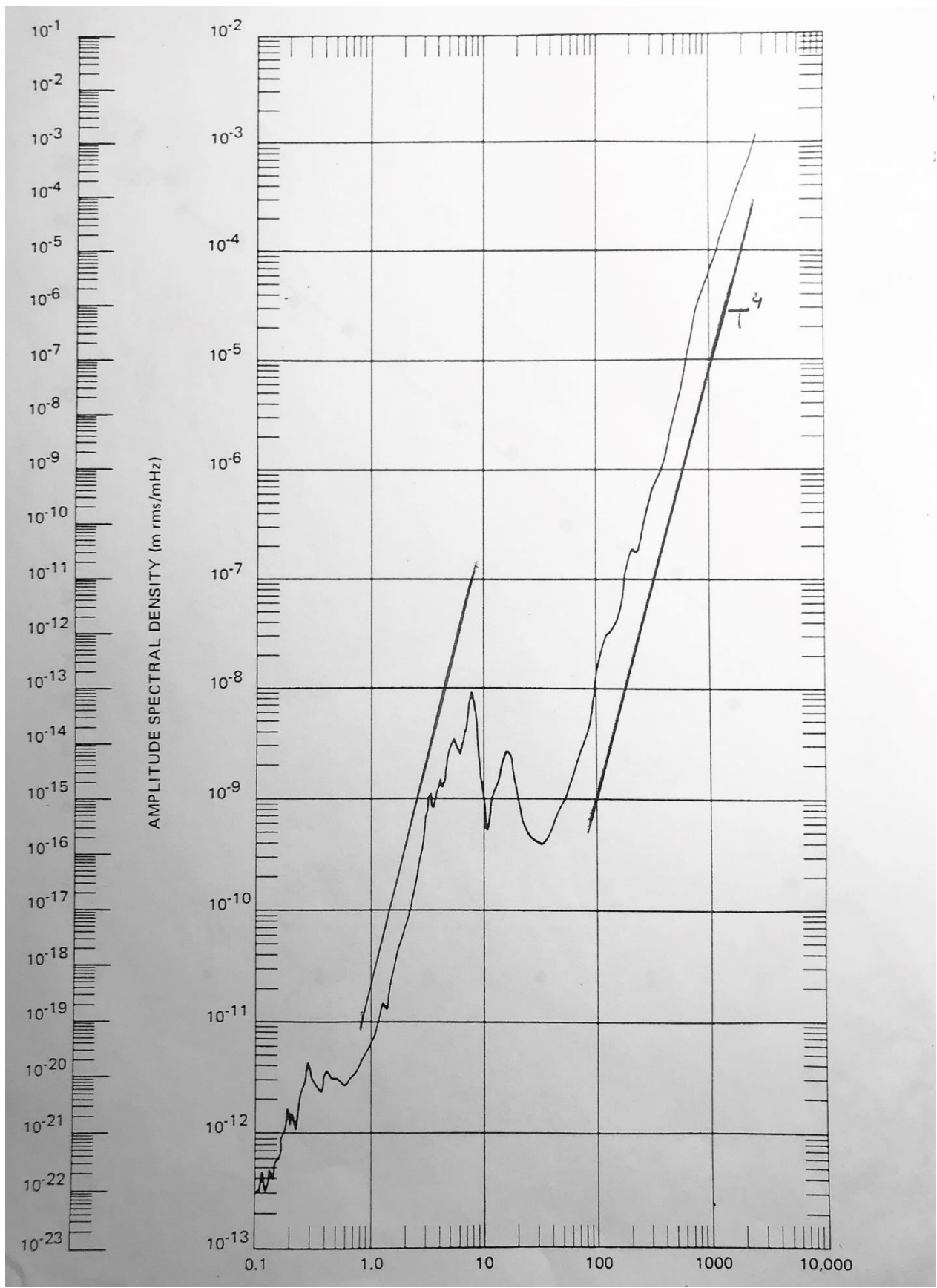

Fig. 2.

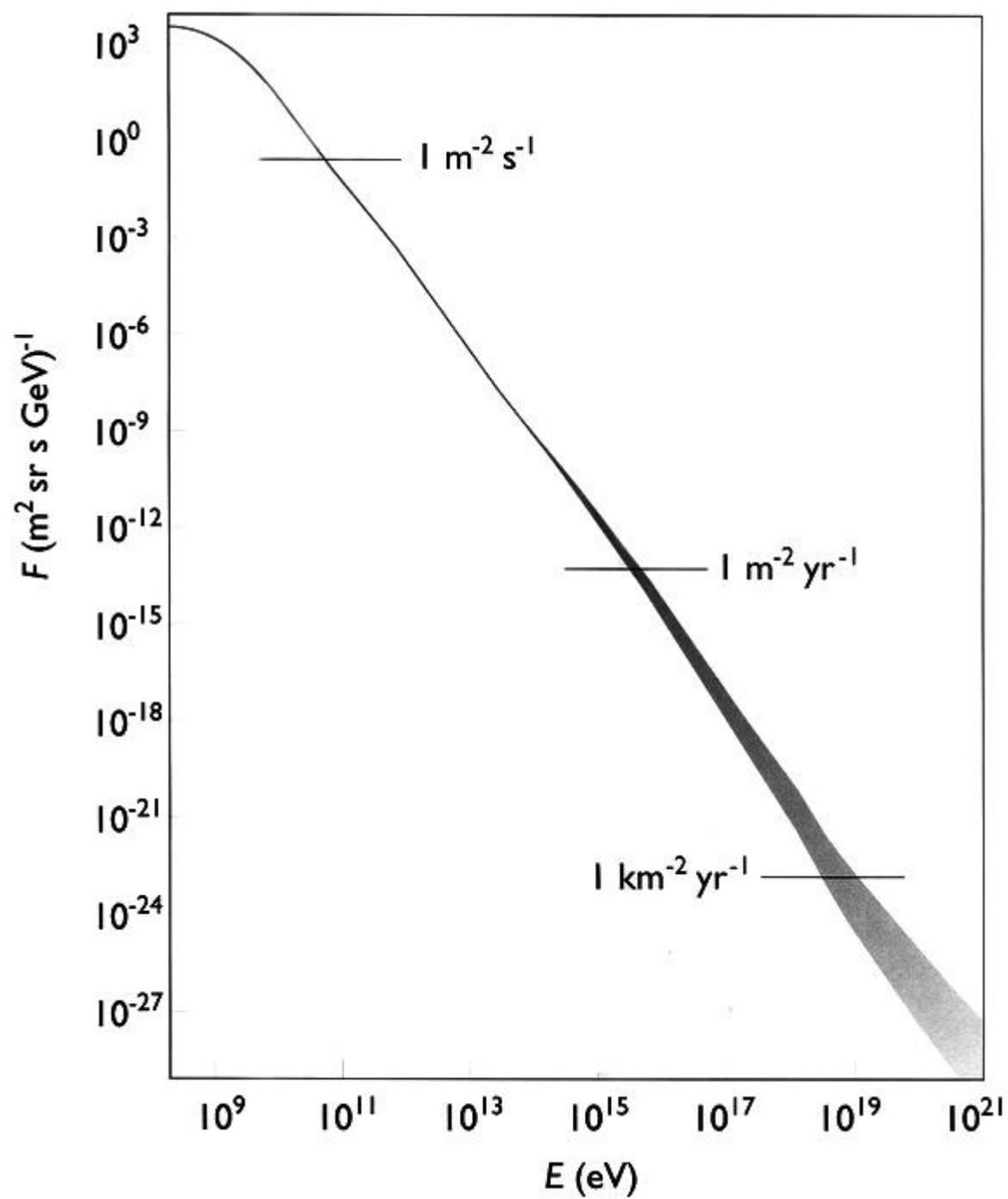

Fig. 3.

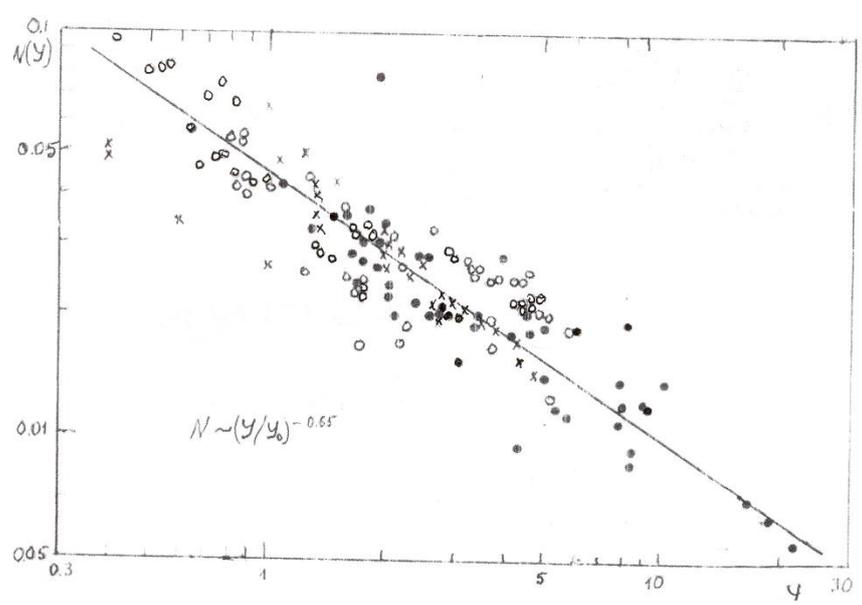

Fig. 4.